# *Unprecedented Severe Atomic Redistribution in Germanium Induced by MeV Self-Irradiation*

Tuan T. Tran and Daniel Primetzhofer

*Department of Physics and Astronomy, Ångström Laboratory, Uppsala University, Box 516, SE-751 20 Uppsala, Sweden*

Corresponding author: Tuan Tran (Tuan.Tran@physics.uu.se)

We present a pronounced unprecedented surface modification of a crystalline Ge layer under heavy ion irradiation with a Ge ion beam at high energy of 2.5 MeV. Under the irradiation conditions, the Ge layer did not become porous as observed for other projectiles and lower energies but develops into an uneven ripple morphology in which the roughness monotonically increases with the irradiation doses. We show that this phenomenon is caused neither by surface erosion effect nor by a non-uniform volumetric expansion. Rather, atomic redistribution in the bulk of the material is the only drive for the ripple surface. Furthermore, the deformation of the Ge layer likely occurs to largest extend after irradiation, as indicated by the very flat interface around the end-of-range region. The observed morphology modification is discussed based on irradiation-induced plastic flow, coupled with a larger contribution of the electronic component in the ion-solid interactions.

Understanding irradiation effects on solid materials by energetic ions is of high technological relevance, for numerous processes such as ion implantation, sputtering, the design of fusion reactor wall and nano-patterning. Particularly under heavy ion irradiation, characteristic post-implant morphologies have been shown to occur for certain materials. Using a low energy $Ar^+$ beam (0.075 − 1.8 keV), Facsko *et al.* have shown that ordered nanometer semiconducting dots are formed on a gallium antimony (100) substrate irradiated at the doses of ~$10^{17}$ − $10^{18}$ cm$^{-2}$ [1,2]. This finding offers an efficient way to produce self-organized quantum dots on semiconducting materials with a single processing step. Comparable phenomenon is also found to occurs for germanium (Ge) irradiated at elevated temperature (350 ℃). A 1 keV $Ar^+$ beam under normal incidence was shown to induce a checkerboard pattern with four-fold symmetry of nanoscale features on Ge surface [3]. Ripple patterns were observed for Ge(100) [4] and Si(100) [5] irradiated with Kr and Xe ions under off-normal incidence at energies ≤ 2 keV. Several models have been suggested for the characteristic morphologies, such as gradient-dependent sputtering and reflection of the primary ions [4,5], reversed epitaxy due to surface diffusion [3] and ion-impact induced prompt atomic redistribution [6].

Fundamentally, energy transfer mechanisms between ions and solids are varied in different projectile energy regimes. At low ion energies (< 10 keV), nuclear interactions are the dominant mechanism. Whereas, energy transfer through electronic interactions prevails in the medium range of energies, such as from tens to hundreds of keV. At higher energies, the projected range of the ions is also substantially longer, prompting a larger contribution of the bulk properties. Therefore, the characteristics of the morphology are different at medium energies. Indeed, under ion irradiation at hundreds of keV, Ge shows different modifications from shallow pits [7] to sponge-like structures [8-10]. Using a 280 keV $^{209}$Bi ion beam, Appleton *et al.* were the first to report a porosity effect in Ge implanted at the dose of ~4 · $10^{15}$ cm$^{-2}$ at room temperature [8,11,12]. The similar effect was later found in many other experiments with different ion species, ion energies, implant doses and substrate temperatures [9,13,14]. In general, for self-implantation at room temperature, i.e. Ge ions into Ge substrates, the material structure will change from crystalline to amorphous at the threshold dose of ~3 · $10^{13}$ cm$^{-2}$, and from amorphous to porous at the threshold dose of ~2 · $10^{15}$ cm$^{-2}$ [14,15]. It is believed that the mobility of defects and the coalescence of the ion-induced vacancies, particularly in the near-surface regions, is the mechanism for the formation of the pores [9,11]. Evidently, low substrate temperatures [9,10] and capping layers [10,16] can reduce the surface diffusion and the nucleation of vacancies on the surface, hence effectively suppressing and delaying the porosity up to the implant dose of $\sim 10^{17}$ cm$^{-2}$ for Ge implants [10] and $\sim 3 \cdot 10^{16}$ cm$^{-2}$ for Sn implants [16].

At even higher ion energies, i.e. in the high energy regime (MeV), much fewer studies on the subject have been reported, which can be partially motivated by decreasing achievable particle doses with increasing energy. Another difference of the regime is the increasing contribution of electronic energy deposition and decreasing nuclear stopping, while at the same time, no continuous track formation is expected for crystalline Ge for monoatomic ions of any kind, also at even higher energies[17]. Using a krypton ion beam at 1.5 MeV at room temperature, Wang *et al.* has shown the formation of sponge-like structures on the surface of Ge, starting at the dose of $\sim 7 \cdot 10^{14}$ cm$^{-2}$ [18]. Comparably, Steinbach *et al.* used an iodine ion beam at 3 MeV to form porosity on the surface and buried void-rich band in Ge [19]. To the best of our knowledge no study employing Ge samples with a capping layer has been reported for the MeV regime. The capping layer might lead to profound differences in the irradiation effects because surface-related contributions are very much suppressed in this case. Reducing a number of the contributing factors, such as the surface diffusion and surface sputtering, might help us to understand the mechanism of the ion-irradiation effects in Ge. In this report, we present an experimental study of the ion irradiation effect on Ge implanted with a broad 2.5 MeV Ge$^+$ ion beam at room temperature. The choice of energy is made to obtain similar specific energy loss but with reversed contribution from electronic excitations and nuclear energy loss when comparing to energies around 500 keV, i.e. electronic stopping exceeding nuclear stopping by about a factor of 2 [20]. The pre-implanted samples were deposited with a capping layer. The post-implanted samples were characterized by cross-section transmission electron microscopy (XTEM), atomic force microscopy (AFM) and Rutherford backscattering spectrometry (RBS).

The initial samples for the experiment were $\sim 0.5$ µm crystalline Ge layer on Si(100) substrates. A capping layer of silicon nitride was deposited onto the sample using reactive sputtering of a silicon target. Different thicknesses of the capping layers were employed ranging from $\sim 10 - 60$ nm to prevent the Ge surface from ion erosion. Rutherford backscattering spectrometry using a 2 MeV He$^+$ beam provided by a 5 MV pelletron tandem accelerator at Uppsala University was employed to check the film quality before implantation (see supplemental material). The ion irradiation of these samples was conducted with a 2.5 MeV Ge$^+$ beam provided by the same accelerator. The samples were kept at room temperature and tilted a few degrees off the beam axis to avoid channelling. Four nominal implant doses were done: $0.7, 1.0, 1.5, 1.7 \cdot 10^{16}$ cm$^{-2}$. The fluence rate during implantation was $0.44 \cdot 10^{12}$ s$^{-1}$cm$^{-1}$, equivalent to a power density of ~0.2 W/cm$^2$. Thus, no significant beam induced heating of the sample is expected.

XTEM was employed to characterize the cross section of the $1.5 \cdot 10^{16}$ cm$^{-2}$ sample. The TEM lamella was prepared using a dual-beam focused ion beam (FIB) system and *in-situ* grid transfer. TEM characterization was conducted in TEM mode at 200 kV with a FEI Titan Themis 200 system. We also employed atomic force microscopy to examine surface topography of the samples. The technique provides intuitive presentation of the surface and precise quantification of the roughness as a function of the implant doses. Finally, large-scale areal density and surface roughness are studied further using broad beam Rutherford backscattering spectrometry operated with a 2 MeV He$^+$ beam. Analysis of the RBS spectra was done by fitting a simulated spectrum to the measurements using the SIMNRA program [21]. For the electronic stopping cross-section, we used the SRIM stopping power database because it is a semi-empirical calculation that relies on experimental values. Hence, SRIM usually provides more reliable data. The SIMNRA simulation program also allows us to incorporate roughness and thickness variations of the thin films and the substrate into the simulation.

In Fig. 1, the TEM micrograph of the $1.5 \cdot 10^{16}$ cm$^{-2}$ sample shows several distinct regions. The darkest top layer is platinum (Pt) deposited on top of the post-implant sample before ion milling to protect the examined region from ion beam damages. Below the Pt layer are the SiN capping and the Ge layers. The Ge layer appears homogenous and free of voids. According to our simulation using the Stopping and Range of Ions in Matter (SRIM) [20], the expected atomic structure of this Ge layer is amorphous (see Supplementary section). The 2.5 MeV Ge$^+$ ion beam induces at least 20 displacements to every Ge atom in the substrate, far more sufficient than the required displacement of 0.3 dpa for complete amorphization [15]. The ion beam is also expected to amorphize the topmost ~1.45 µm of the Si substrate as apparent from a rather sharp buried interface visible in Fig. 1. Most remarkably, however, the Ge here appears to show an irregular and pronounced surface roughness with an amplitude of almost half the film thickness. The average thickness of the Ge film seems to be in good agreement with the nominal thickness and the SiN film appears to be intact covering all Ge.

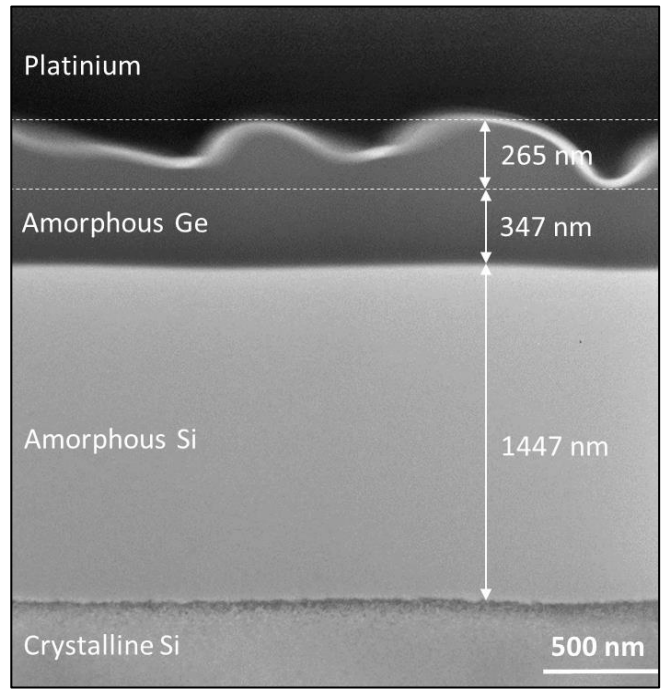

Fig. 1: Cross sectional transmission electron micrograph of the Ge-on-Si substrate after irradiation with 2.5 MeV Ge+ ions at a dose of $1.5 \cdot 10^{16}$ cm$^{-2}$.

Further characterization of the surface topography was done on all implanted samples using AFM as shown in Fig. 2. Starting with a flat surface with the root-mean-square (RMS) roughness $R_q \approx 1$ nm, the surface has already become rippled at the lowest dose of $0.7 \cdot 10^{16}$ cm$^{-2}$ (Fig. 2a). The amplitude and the sizes of the ripples apparently increase monotonically with the implant doses (Fig. 2b). The RMS roughness of the highest dose samples is ~50 nm.

We will in the following consider three possible mechanisms to account for the observation. Firstly, surface roughness can be induced by ion sputtering. However, this explanation would require a material removal rate well above expectations for the employed doses. Moreover, the capping layer, which is seemingly only little affected, is expected to prevent the material from sputtering. Secondly, the ripple surface might be an irradiation induced fluctuation of the mass-density within the Ge layer after irradiation – however, the apparent thickness from the TEM-micrograph is also rendering this explanation unlikely. Finally, we might observe an uneven redistribution of the Ge atoms during or after implantation.

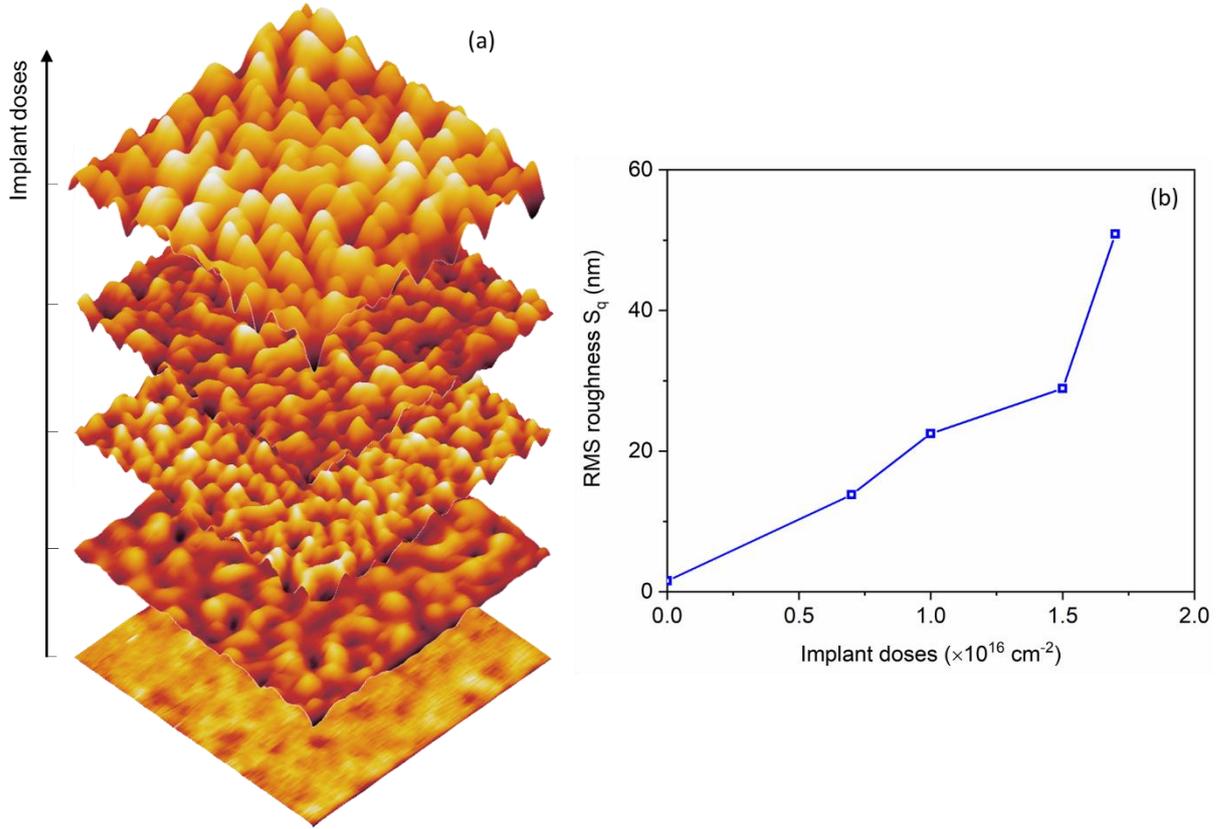

Fig. 2: (a) Surface topography of the as-irradiated samples acquired by an atomic force microscope, and (b) root-mean-square roughness of the surface as a function of the implant doses.

In Fig. 3, we show the RBS spectra measured for all samples (data points) and the corresponding SIMNRA simulation (solid lines). The high-energy edges of the Ge peaks are located at different energy because the thickness of the SiN capping layers is different for the samples. All high-energy edges are found to be identical in slope, which indicates a well-defined thickness of the SiN-layers which furthermore remains almost unaffected by the irradiation. Noticeably, the low-energy edges of the peaks are deteriorating for increasing irradiation dose. The uneven redistribution of the Ge atoms is found to be more severe at higher implant dose as indicated by the decreasing slope of the low-energy edges. At the same time, the area of the Ge signal remains unaffected as the integrated yields of the Ge peak are almost identical for all samples. Therefore, it is affirmative that sputtering effect can be ignored and do not play any role in the roughness phenomenon.

Revisiting the TEM micrograph of Fig. 1, we have found the total area of the cross section of the a-Ge layer $1.773 \cdot 10^6$ nm$^2$. According to the SIMNRA simulation, the thickness of Ge layer is 496.8 nm. Therefore, the total area of the Ge layer having a width equivalent to the one in Fig. 1 is $1.738 \cdot 10^6$ nm$^2$. The resulting volumetric difference between the pristine and the sample implanted with the dose of $1.5 \cdot 10^{16}$ cm$^{-2}$ is thus less than 2%, confirming our refusal of the density fluctuations. The fluctuation of the mass density would also have led to pronounced contrast in the bright field TEM image in Fig. 1. Hence, we conclude that the roughness of the surface is thus solely induced by an uneven redistribution of the Ge atoms caused by the implantation.

In Fig. 1 we also note that the interface between the amorphous Si layer and the crystalline Si substrate are surprisingly flat. Its morphology does not follow that of the a-Ge layer as expected due to the constant end-of-range of the ion beam. This feature strongly suggests that the deformation of the a-Ge layer happened after the implantation, when the beam was off. This phenomenon is probably caused by the on/off effect as reported by Volkert [22], such that when the beam is on the material is densified. When the beam is off, the stress starts to increase due to the expansion of the irradiated area.

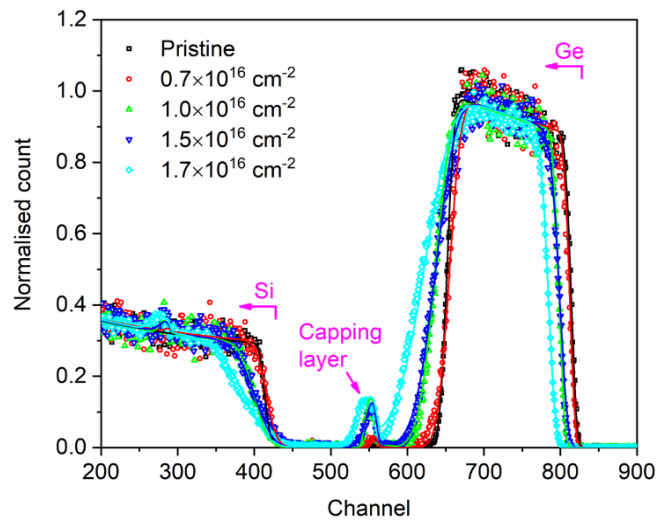

Fig. 3: Rutherford backscattering spectrum of the Ge-on-Si substrates after irradiation with different doses, from $0.7 - 1.7 \cdot 10^{16}$ cm$^{-2}$. Scattered points and solid lines represent measurements and corresponding SIMNRA simulations, respectively.

Noticeably, the Ge layer in this study does not develop into a porous structure as observed for an implant at this dose and ion energies of several hundred keV. We have mentioned that porosity in Ge consistently occurs for the implant dose $> 2 - 3 \cdot 10^{15}$ cm$^{-2}$ at room temperature for the medium energy regime [10,13,14] and for the high energy regime [18,19]. Even with a capping layer, Ge becomes porous at the dose of $> 5 \cdot 10^{15}$ cm$^{-2}$ [10]. Following is our discussion on the profound difference in the morphology between the samples of this study and the preceding.

It is known that irradiation leads to lattice stress due to the creation of defects in the lattice [22,23]. Irradiating thick crystalline Ge with a 3 MeV iodine ion beam at room temperature, Steinbach *et al.* showed that the mechanical stress peaked at the dose of $1 \cdot 10^{13}$ cm$^{-2}$ and then saturated at higher doses [23]. Volkert *et al.* also reported the similar effect for crystalline and amorphous Si, silicon germanium alloys, amorphous silicon dioxide and polycrystalline aluminum films [22,24]. A common factor of studies reporting porosity [10,14,18,19] is that the specific nuclear energy deposition is equivalent or even higher than the electronic energy loss [20]. For example, for a 140 keV Ge$^+$ beam irradiated onto Ge substrates the electronic and the nuclear losses are 20.4 eV/Å and 151.8 eV/Å, respectively. In contrast, for a 2500 keV Ge$^+$ beam the electronic and the nuclear losses are 117.7 eV/Å and 55.6 eV/Å, respectively. Based on this, one can understand, that in our system, the excitation of the material at the nanoscale looks significantly different, with less violent collision cascades, but in comparison high local electron temperatures which is via electron-phonon coupling dissipated in the lattice. We speculate, that this effect can strongly affect the mobility of Ge recoils in our film, in particular at the low-energy tail of the collision cascade, and while stress can accumulate in the system, prevents formation of voids. Subsequently, and to significant extend post-irradiation plastic flow occurs as a mechanism for stress relaxation and is expected to be the main driving force for the atom redistribution observed in this study.

In conclusion, we study irradiation effects on a crystalline Ge layers induced by self-irradiation with a 2.5 MeV Ge ion beam at room temperature. The data shows severe surface modification in all implanted samples as the surface roughness increases with the implant doses. Furthermore, our measurements indicate that the significant deformation of the Ge layer happens at least partially only after the irradiation, as presented by the very flat interface around the end-of-range of the ions. The observed modification of the germanium layer is not caused by preferential ion sputtering observed in alloy or polycrystalline materials. Neither, volumetric expansion nor fluctuation of mass density within the Ge layer are the origin of the phenomena. We have shown that inhomogeneous redistribution of the Ge atoms during the irradiation is the sole cause of the severe roughness of the sample surface. The driving force of the atomic redistribution is expected to be irradiation-induced stress and subsequent plastic flow, coupled with a larger impact of the electronic component of the ion-solid interactions.

## Acknowledgement

Support by VR-RFI (contracts #821-2012-5144 & #2017-00646_9) and the Swedish Foundation for Strategic Research (SSF, contract RIF14-0053 and SE13-0333) supporting accelerator operation is gratefully acknowledged.

# Supplementary

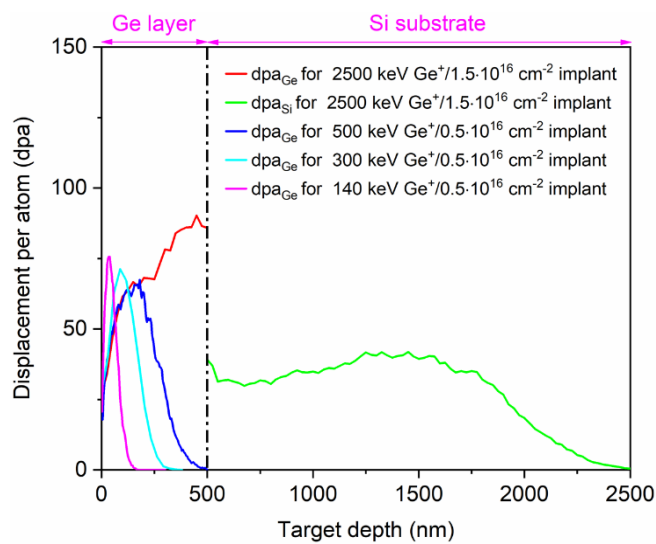

Supp. Fig. 1: Number of displacements per atom as calculated by SRIM for 140 keV and 2500 keV Ge+ implants. The calculation was done in full cascade mode in which both primary ion paths and all subsequent recoil paths are followed [20]. Despite having comparable dpa as the 140 keV Ge+ implant, the 2500 keV Ge+ implant does not create a porous structure, but a severe atomic redistribution, suggesting a larger role of electronic interactions in the atomic redistribution.

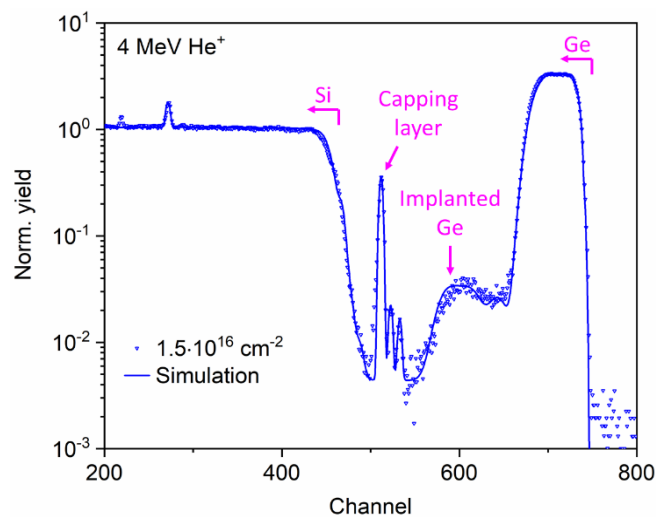

Supp. Fig. 2: RBS spectrum of sample irradiated with a dose of $1.5 \cdot 10^{16}$ cm$^{-2}$ recorded with a 4 MeV He+ beam to resolve the implanted Ge atoms in the Si substrate. The measured implant dose is close to the nominal value.

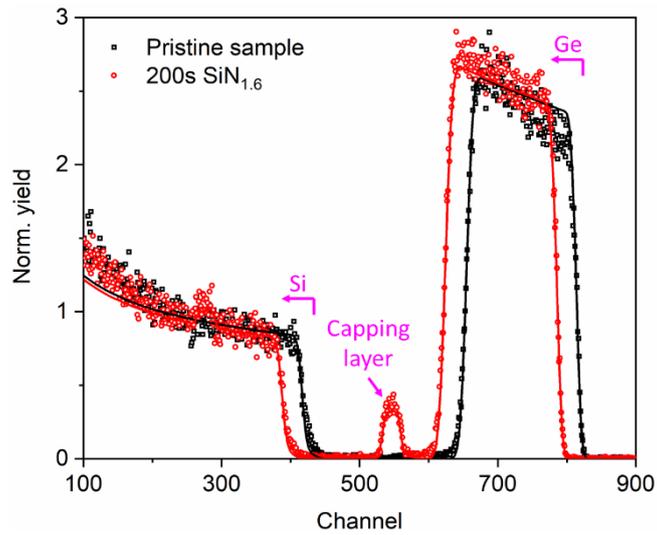

Supp. Fig. 3: RBS spectra of the pristine Ge-on-Si sample (black) and the sample after 200s of SiN deposition (red). The data points and the solid line represent the measurement and the SIMNRA simulation, respectively. Based on the slope of the front-edge and the back-edge of the Ge peaks, it is shown that the Ge layer is not affected by the deposition process. Within the measurement error of 3%, the areal density of the two samples is virtually equal. Hence, neither sputtering nor a significant morphological change of the Ge layer occurred during the process.